\documentclass[doublecol,figures]{epl2}

\usepackage{amsmath,amssymb}
\usepackage{multirow}

\title{Towards an objective ranking in online reputation systems: the effect of the rating projection}

\author{Hao Liao$^1$ \and An Zeng$^2$ \footnote{E-mail: anzeng@bnu.edu.cn}\and Yi-Cheng Zhang$^1$}
\shortauthor{Hao Liao \etal}
\institute{
  \inst{1}D{\'e}partement de Physique, Universit{\'e} de Fribourg - Chemin du Mus{\'e}e 3, CH-1700~Fribourg, Switzerland\\
  \inst{2}School of Systems Science, Beijing Normal University, Beijing 100875, P. R. China
}

\pacs{89.65.-s}{Social and economic systems}
\pacs{89.20.Ff}{Computer science and technology}
\pacs{89.20.Hh}{World Wide Web, Internet}

\abstract{Online reputation systems are commonly used by e-commerce providers nowadays. In order to generate an objective
ranking of online items' quality according to users' ratings, many sophisticated algorithms have been proposed in the literature.
In this paper, instead of proposing new algorithms we focus on a more fundamental problem: the rating projection. The basic idea is that even though the rating values given by users are linearly separated, the real preference of users to items between different values gave is nonlinear. We thus design an approach to project the original ratings of users to more representative values. This approach can be regarded as a data pretreatment method. Simulation in both artificial and real networks shows that the performance of the ranking algorithms can be improved when the projected ratings are used.}
\date{\today}

\begin{document}

\maketitle

\section{Introduction}
The coming big data era brings us an critical problem: how to extract the valuable information from the big data at hand. This problem is especially crucial in online systems where the available data are overwhelmingly abundant due to the rapid expansion of the Internet~\cite{WATTS,ALBERT,GUTTMAN,KLEINBERG}. To filter out irrelevant online items (e.g. books, movies or others) for users, the recommender system, such like the collaborative filtering methods are widely applied \cite{CF1,CF2}. Besides the relevance, the quality of items is also of great importance to online users. Therefore, many online websites, such as \emph{Amazon.com} and \emph{Netflix.com} build the online reputation system \cite{repu1,Masum,repu2,repu3} in which users can give their opinions to an item by assigning certain rating value to it. The purpose of the reputation system is to help users uncover the true quality of items. After obtaining the rating data, some algorithms are needed to generate the ranking of items. The most straightforward way is to simply use the arithmetic average of ratings to rank items' quality. However, since this method has low ranking accuracy and is sensitive to spamming behavior, many other ranking algorithms have been proposed recently \cite{An}.

Other types of ranking algorithms compute users' reputation and items' quality self-consistently. More specifically, these algorithms usually update users' reputation in an iterative way and aggregate the ratings based on the reputation of users \cite{YU}. A representative one of these algorithms is called iterative refinement (IR) \cite{EPL06}. In IR, a user's reputation is inversely proportional to the mean difference between his rating vector and objects' estimated quality vector (i.e., weighted average rating based on user reputation). The estimated quality of objects and reputation of users are iteratively updated until the values reach a stationary point. This method is further modified by assigning trust to each individual rating \cite{Mizz ,EPL10}. Recently, Zhou et al. \cite{EPL11} takes the robustness of the algorithm into account and propose to calculate a user's reputation by the Pearson correlation \cite{correlation} between his ratings and objects' estimated quality. This method is usually referred to as the Correlation-based Ranking (CR) method and it can be resistent to the malicious spamming behaviors of some users.

However, a fundament problem in the reputation system has been neglected for a long time. For most online reputation systems, the rating values are discrete and linearly separated. For example, some well-known websites such as \emph{Amazon.com} and \emph{Netflix.com} use the 5-star rating system: users are allowed to rate items with integers from 1 (worst) to 5 (best) \cite{Jøsang,PAN}. However, the real preference of users to items between different rating values can be actually nonlinear. For instance, the difference between ratings 4 and 3 might not be equal to that between 5 and 4. Based on this idea, we design a rating projection method which allows us to project the original rating values to more representative ones. We consider both artificial and real networks. The projected ratings are then used as input to several ranking algorithms and significant improvement in the ranking accuracy is observed.

\section{Rating projection method}
The reputation systems can be normally described by weighted bipartite networks consisting of online users and items. If a user rated an item, there is a link between them, and the link weight is the rating value that the user gives to the item. Here, we consider a common case where users rate items by using the integer scale from 1 (worst) to 5 (best). In this 5-star rating system, 3 means neutral. However, when a user rates item $\alpha$ with 4 and item $\beta$ with 5, it doesn't mean that the user like $\beta$ two times more than $\alpha$. This problem also exists when one compares rating 1 and 2. Based on this idea, we design a rating projection method. Since rating 1, 3 and 5 respectively stand for the worst, neutral and best, we preserve these three ratings. The method transfers the rating values 2 and 4 to new values $R_2$ and $R_4$ via
\begin{equation}
R_2= 1 + p_1 * 2
\end{equation}
\begin{equation}
R_4= 3 + p_2 * 2,
\end{equation}
where $p_1$ and $p_2$ are tunable parameters. Clearly, when $p_1=0.5$ and $p_2=0.5$ are assigned, it gives the original rating values. However, the rating values becomes nonlinear when $p_1\neq0.5$ and $p_2\neq0.5$. In the following, we will investigate the performance of different ranking algorithms when the projected ratings are used.

\section{Ranking algorithms}
In this work, we mainly consider four ranking algorithms in our experiments. We first introduce some notations for these ranking algorithms. The users are denoted by set $U$ and items are denoted by set $O$. To better distinguish different types of nodes in the bipartite network, we use Latin letters for users and Greek letters for items. The rating given by a user $i$ to an item $\alpha$ is denoted by $r_{i\alpha}$. Moreover, we define the set of items selected by user $i$ as $O_i$ and the set of users selecting item $\alpha$ as $U_{\alpha}$, and the degree of users and objects are respectively $k_i$ and $k_{\alpha}$. We also denote quality of item $\alpha$ and the reputation of user $i$ as $Q_{\alpha}$ and $R_i$, respectively.

(i) The first algorithm is the so-called \emph{mean} method. In this method, the quality of an item is simply the mean ratings the item received. As this method is easy to calculate, it is widely used by many websites. However, as this method doesn't consider user's reputation, it is sensitive to malicious manipulations.

(ii) The \emph{iterative refinement} (IR) calculates user reputation and item quality in a self-consistent way. It considers a user's reputation as inversely proportional to the mean squared error between his/her rating vector and the corresponding objects' weighted average rating vector \cite{EPL06}. The estimated object quality values $Q_{\alpha}$ is defined as
\begin{equation}
Q_{\alpha}= \frac{\sum_{i\in{U_{\alpha}}}{R_ir_{i\alpha}}}{\sum_{i\in{U_{\alpha}}}{R_i}},
\end{equation}
and the estimated reputation of user $R_i$ is computed as
\begin{equation}
  R_{i} = \left({\frac{1}{\lvert O_{i} \rvert} \sum_{\alpha\in {O_{i}}} (r_{i\alpha} - Q_{\alpha})^2} + \varepsilon\right)^{-\beta},
\end{equation}
where $\beta$ is a tunable parameter. Note that IR will simply equal to the Mean method when $\beta=0$. Here we set $\beta=1$ because IR performs best under this setting \cite{EPL11}. The algorithm is initialized by setting the reputation $R_{i}=1$ for all users $i$ and stops when $Q_{\alpha}$ and $R_i$ reached a stationary point in iterations.

(iii) The remaining two methods, namely \emph{Correlation-based ranking} (CR) and \emph{Reputation redistribution ranking} (RR), are actually based on the same framework. Here, we mainly discuss the RR method. The initial configuration of RR for each user is set as $R_i = {k_i}/\lvert O \rvert$. The quality of an object depends on users' rating and can be calculated by the weighted average of rating to this object. Users' estimated reputation $R_i$ and items' estimated quality $Q_{\alpha}$ are updated in the following way,
\begin{equation}
Q_{\alpha}= F\cdot\frac{\sum_{i\in{U_{\alpha}}}{R_ir_{i\alpha}}}{\sum_{i\in{U_{\alpha}}}{R_i}},
\end{equation}
where $F = \max_{i\in{U_{\alpha}}}\{R_i\}$ is an penalty factor eliminating the effect of an object which is rated with a high score only by one or two users.

Before obtaining $R_i$, one has to first calculates a quantity called temporal reputation $TR_i$ as
\begin{equation}
TR_i= G \cdot\frac{1}{k_{i}}\sum_{\alpha\in{O_i}}{(\frac{r_{i\alpha}-\bar{r_i}}{\sigma_{r_i}})}{(\frac{Q_{\alpha}-\bar{Q_i}}{\sigma_{Q_i}})}.
\end{equation}
where $G = {lg(k_i)}/{\max\{lg(k_j)\}}$ helps the iterative process to filter out the influence of the not yet reliable users.

$TR_i$ is then nonlinearly redistributed to all users via
\begin{equation}
R_i=TR_i^{\theta}\frac{\sum_j TR_j}{\sum_j TR_j^{\theta}},
\end{equation}
where $\theta$ is a tunable parameter which is optimal for $\theta^*= 5$ \cite{An}.

When $F$ and $G$ are absent and $\theta=1$, the RR algorithm degenerates become equivalent to the CR method. In both CR and RR algorithms, users' reputation and items' quality are updated in each step. The iteration stops when $|{Q-Q'}|= ({1/\lvert O \rvert})\sum_{l\in{O}}{(Q_l-{Q_l}')^2}$ is smaller than the threshold value $\Delta = 10^{-4}$. Here $Q'$ denotes the quality value at the previous iteration step. In general, CR can outperform the mean and IR methods, but RR can create a more accurate and robust ranking than CR, especially when the online systems have many spammers.

\section{Results on real networks}
In this section, we select two commonly used real data sets in online rating systems: Netflix and MovieLens. MovieLens is provided by GroupLens project at University of Minnesota \footnote{www.grouplens.org}. We use a subset of the complete data. In the subset, 1 million ratings are given by $6040$ users to $3706$ items based on the integer rating scale from 1 to 5.  Netflix is a huge data set released by the DVD rental company Netflix for its Netflix Prize \footnote{www.netflixprize.com}. We extracted a smaller data set by randomly choosing $5000$ users and took all $16195$ movies they had rated. There are finally $1070000$ ratings in the Netflix are also based on 5-star system. Some basic characteristics of these data sets are summarized in table~\ref{tab1}. 
\begin{table}
  \centering
  \caption{Some basic characteristics of the real data sets considered in this paper.}\label{tab1}
  \begin{tabular}{l| cccc cccc cccc cccc cccc }
    \hline\hline
Methods & $|U|$ & $|O|$ & $\langle k_u \rangle$ & $\langle k_o \rangle$ & Sparsity \\ \hline
MovieLens &6040	&3706	&166	&270	&0.0447\\
Netflix &5000	&16195	&214	&66	&0.0132\\
    \hline\hline
  \end{tabular}
\end{table}

To test the ranking performance, we use a standard accuracy measure called ranking score (RS) index \cite{PR}. We first select a group of high quality benchmark items $E$ which are nominated at Annual Academy Award ($203$ in Movielens and $293$ in Netflix). For each of these awarded items $\alpha$, its rank is denoted as $D_\alpha$ among all $M$ items. Then RS can be calculated as
\begin{equation}
\label{eq:RS}
RS=\frac{1}{|E|}\sum_{\alpha\in E} \frac{D_\alpha}{M}.
\end{equation}
According to the definition, an accurate ranking should have a small RS.

The heatmap of $RS$ in the parameter space $(p1,p2)$ is shown in Fig. 1. One can immediately notice that $p_2$ dominates the performance of the rating projection method. More specifically, a better $RS$ will be achieved when $p_2<0.5$ in Movielens and $p_2>0.5$ in Netflix. These results are consistent when CR and RR are applied. It suggests that in Movielens, when users give rating 4 to a movie this rating is closer to 3. In Netflix, when users rate a movie with 4, this rating is closer to 5. On the other hand, the influence of $p_1$ on RS is minor. When $p_1$ is larger than $0.5$, RS is slightly improved. The optimal $p_1^*$ and $p_2^*$ in these two real systems are respectively ($0.75$, $0.25$) and ($0.85$, $1$).

\begin{table*}
  \centering
  \caption{RS values of different algorithms and rating projection based algorithms for the real-data sets.}\label{tab2}
\begin{tabular}{|c|c|c|c|c|c|c|c|c|}
\hline
\multirow{2}{*}{Method} &
\multicolumn{2}{c|}{Mean} &
\multicolumn{2}{c|}{IR} &
\multicolumn{2}{c|}{CR} &
\multicolumn{2}{c|}{RR} \\
\cline{2-9}
  & Original & Projected & Original & Projected & Original & Projected & Original & Projected  \\
\hline
MovieLens & 0.127 & \textbf{0.124} & 0.124 & \textbf{0.122} &  0.128 & \textbf{0.125} & 0.098 & \textbf{0.092} \\
Netflix & 0.271 & \textbf{0.258} & 0.254 & \textbf{0.245} & 0.254 & \textbf{0.244} & 0.114 & \textbf{0.102} \\
\hline
\end{tabular}
\end{table*}

\begin{figure}
\centering
\includegraphics[width=0.49\textwidth]{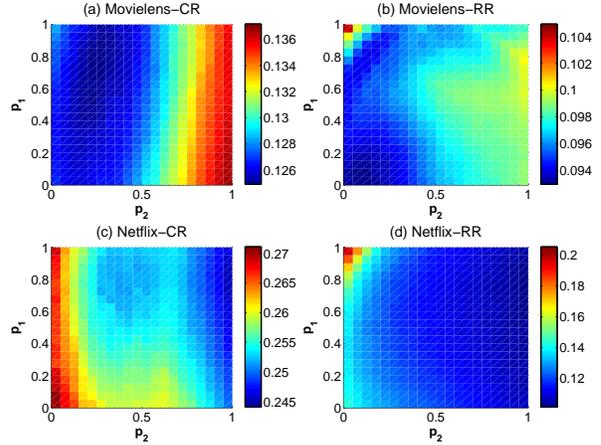}
\caption{RS of different methods in the parameter space $(p_1,p_2)$.
}
\label{fig:scale}
\end{figure}

To show the detailed improvement from the rating projection method, we report in table 2 the $RS$ value of different ranking methods based on the original ratings and optimal projected ratings. One can see that the $RS$ is indeed improved when the optimal projected ratings are used. In Movielens, the improvement reaches to $2\%$ in the Mean method and $2\%$ in the CR method, $2\%$ in the IR method, and $6\%$ in the RR method. In Netflix, the improvement reaches to $5\%$ in the Mean method, $4\%$ in the CR method, $4\%$ in the IR method, and $11\%$ in the RR method. Among these four algorithms, the best one, as we mentioned above, is RR. The remarkable advantage of rating projection is that its improvement to RR is most significant among these methods. Our results on real data also suggest that the projection of rating 4 is crucial. However, how this rating value should be adjusted depends on different real systems.


\section{Result on artificial data}
\subsection{Generating artificial networks}
In order to understand why the rating projection method improves the ranking algorithms in-depth, we construct the artificial networks which mimics users behavior in real online rating system. We set number of users to $|U|=6000$ and item number $|O|= 4000$ in the artificial network. To generate the user-object bipartite network, the links are added to the network one by one until the network reaches a target sparsity ($\phi = 0.2$). Under this setting, the final network will have $\phi|U||O|=4.8*10^5$ links. As preferential attachment widely occurs in many real systems, we add links to the network according to it. Specifically, the probabilities for selecting a user $i$ and object $\alpha$ to create a new link is proportional to the degree they already have~\cite{BARABASI}.

After creating the network structure, the link weight (rating) needs to be determined. We assume that each object $\alpha$ has a real intrinsic quality denoted by $Q'_{\alpha}$. When a user $i$ gives a rating to the object $\alpha$, he/she will inevitably have personal error $e_{i\alpha}$ which is determined by this user's magnitude of rating error $e_{i}$. Accordingly, the object $\alpha$ quality estimated by user $i$ will be
\begin{equation}
q_{i\alpha}=Q'_{\alpha} + e_{i\alpha}.
\end{equation}
In our simulation, considering most of real online rating systems have the rating bounds $[1,5]$, our artificial data keep a constant rating limits $[1,5]$. For each object, the intrinsic quality $Q'_{\alpha}$ will be drawn from a uniform distribution $[1,5]$, and personal error magnitude $e_{i\alpha}$ is drawn from a normal distribution $(0, e_i)$. For each user $i$, $e_i$ is generated from a uniform distribution $(\delta_{min} = 0, \delta_{max}=4)$. We first adopt a method to get discrete rating value by $r_{i\alpha}=[q_{i\alpha}]$, i.e. set $r_{i\alpha}$ to the nearest integer~\cite{EPL10}. In this case, rating values fall outside the rating range $[1;5]$ are truncated, where those smaller than $1$ are changed to $1$, and those greater than $5$ are changed to $5$. This is a realistic constraint that, no matter how much a user likes or dislikes a particular object, they only can rate it in the given rating bounds. This is considered as the standard method to get discrete ratings and is denoted as case 0 in this paper.

In real rating systems, when users have difficulty in estimating to what extend they like or dislike an item, they may get confused when choosing a integer value for the product from discrete rating bound $[1, 5]$. Therefore, their ratings to the item might not be able to reflect their true preference to the item. In our artificial networks, we consider some other ways to get discrete ratings. This will help us to understand better the phenomenon we observed in real systems.
\begin{enumerate}
\item case 1, we assume the users cannot distinguish the raw ratings within [0, 2.5], so half of the raw rating are randomly selected and set to be 1 and the rest are set to be $2$.
\item case 2, we assume that users cannot distinguish the raw rating in range [1.5, 3.5]. We randomly choose $50\%$ of those raw ratings within this range and set them to be $2$ and the other $50\%$ are set to be $3$.
\item case 3, half of the raw ratings in [2.5; 4.5] are set as 3 and the other half are set as 4.
\item case 4, half of the raw ratings in [3.5, 5] are set as 4 and the other half are set as 5.
\end{enumerate}
The case 1 and 2 consider the situation where users have difficulty in estimating to what extend they dislike the item. The case 3 and 4 consider the situation where users have difficulty in estimating to what extend they like the item. In all the 4 cases above, the raw rating out of the range we mentioned will still follow $r_{i\alpha}=[q_{i\alpha}]$ (the nearest integer rule). We will apply our rating projection method with different algorithms to estimate user reputation and item quality in these 5 cases and study how these five cases influence the results.


\subsection{Results}

For a good algorithm, the estimated user reputation $R_i$ should be negatively correlated with $e_i$ \cite{An}. To quantify the correlation, we calculate the Pearson correlation between $R_i$ and $e_i$ on artificial datasets with different settings~\cite{correlation}. The results reported in Fig. 2. To better show the results, we pick $p_2 = 0.5$ and $p_1 =0.5$ respectively, and plot correlation as function of $p_1$ and $p_2$ in the CR algorithm on these $5$ cases in Fig. 2(a) and (b). As we can see in the case 0, the optimal $p_1$ and $p_2$ is $0.5$, which means that taking the original rating data as the input data for the ranking algorithms is the best. The optimal Pearson correlation for case 1 is achieved when $p_1$ is small. For case 2, the optimal Pearson correlation is achieved when $p_1$ is large. However, for case 3 and case 4, the Pearson coefficient is barely influenced by $p_1$ and $p_2$. These results indicate that when users cannot distinguish to what extend they don't like a product, adjusting $p_1$ can improve the performance of the ranking algorithms. However, if the users don't know to what extend they like a product, adjusting $p_1$ has no influence on the ranking performance. In Fig. 2(b), the influence of $p_2$ on the Pearson correlation coefficient is shown. Different from the results in Fig. 2(a), increasing $p_2$ will decreases the Pearson coefficient for most cases, but only increases the Pearson coefficient for case 3. These results indicate that when users like an item but randomly give ratings between 3 and 5, tuning $p_2$ can improve the ranking performance.

\begin{figure}
\centering
\includegraphics[width=0.49\textwidth]{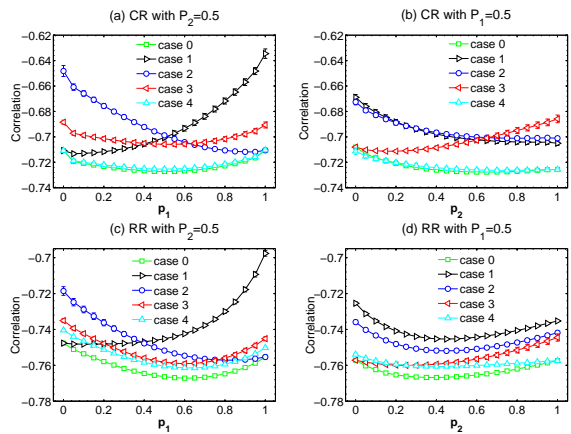}
\caption{
  (colour online) the dependence of the $Correlation$ on $ p_1$  and $p_2$ in CR and RR methods. The results in this figures are averaged over 10 independent realizations. The error bars are the corresponding standard deviations. $p_1$ and $p_2$ corresponding to rating projection method parameter.
}
\label{Fig. 1}
\vspace{-0.2cm}
\end{figure}

\begin{figure}
\centering
\includegraphics[width=0.49\textwidth]{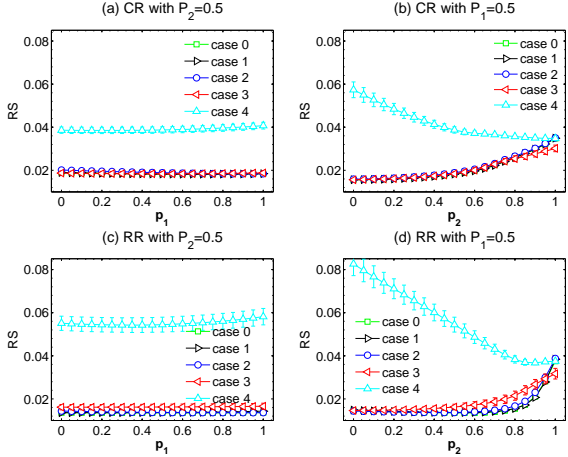}
\caption{
  (colour online) the dependence of the $RS$ on $ p_1$  and $p_2$ in CR and RR methods. The results in this figures are averaged over 10 independent realizations. The error bars are the corresponding standard deviations. $p_1$ and $p_2$ corresponding to rating projection method parameter.
}
\label{fig:scale}
\end{figure}

Besides the CR method, we also investigate the RR method in Fig. 2 (c) and (d). Similarly, $p_1$ mainly affects the case 1 and 2. However, for RR, the effect of $p_2$ to all cases are similar. This is maybe due to the nonlinearity of the penalty factors in RR methods. One interesting observation here is that the pearson correlation coefficient in RR is higher than that in CR, indicating that RR is more effective than CR in uncovering users' reputation.

After investigating the performance of the methods in the user side, we now move to study these methods' ranking accuracy on items' quality. As discussed above, we here use the ranking score metric. To this end, we first select a group of benchmark items $E$ which are top $5\%$ intrinsic quality $Q'_{\alpha}$ among all objects, and compute the ranking score metric based on these benchmark items. The results are reported in Fig. 3. One immediate observation here is that $p_1$ has almost no influence at all for $RS$, as shown in Fig. 3(a) and (c). On the contrary, in both CR and RR methods, $p_2$ can significantly change the $RS$. Selecting $p_2$ here is actually very important. In Fig. 3(b) and (d), one can see that the effect of $p_2$ on RS in case 3 and case 4 is opposite. More specifically, in case 3 a small $p_2$ can lead to a low RS, while in case 4 a large $p_2$ results in a low RS. The results in Fig. 3 also show that in the case 4 where users mess up the ratings between $4$ and $5$ is most harmful for the RS. In this case, tuning $p_2$ can lower the RS, but the RS will still be higher than other cases.

\begin{figure}
\centering
\includegraphics[width=0.49\textwidth]{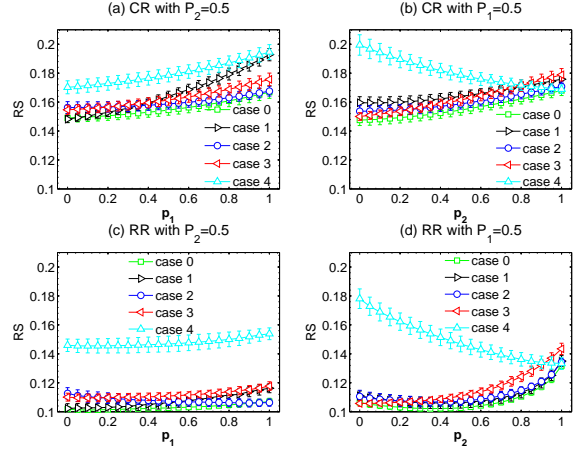}
\caption{
  (colour online) the dependence of the $RS$ on $ p_1$  and $p_2$ in CR and RR methods on spammer data. The results in this figures are averaged over 10 independent realizations. The error bars are the corresponding standard deviations. $p_1$ and $p_2$ corresponding to rating projection method parameter.
}
\label{fig:scale}
\end{figure}

In the real rating system, it is generally difficult to estimate the quality of an object when spammer ratings occured in the dataset. In literature~\cite{An}, it has been shown that spamming problem can be well addressed in RR method. In order to better understand the rating projection method which can be also robust to spamming, we further study the effect of spamming information on the performance of the rating projection method. In practice, we choose random ratings for our artificial method. We first generate the artificial networks according to the rules described above. In order to add some noisy information to the systems, we randomly pick $p$ fraction of the links and replace the rating on each of these links by a random value in range of $[1,5]$. The noisy information in the system increases with the parameter $p$. When $p=1$, there is no original information in the rating system. In the following analysis, we set $p=0.9$.

\begin{table*}
  \centering
  \caption{RS values of different algorithms and rating projection based algorithms for the aritificial-data sets.}\label{tab2}
\begin{tabular}{|c|c|c|c|c|c|c|c|c|}
\hline
\multirow{2}{*}{Method} &
\multicolumn{2}{c|}{Mean} &
\multicolumn{2}{c|}{IR} &
\multicolumn{2}{c|}{CR} &
\multicolumn{2}{c|}{RR} \\
\cline{2-9}
  & Original & Projected & Original & Projected & Original & Projected & Original & Projected  \\
\hline
case0 & 0.127 & \textbf{0.124} & 0.124 & \textbf{0.122} &  0.128 & \textbf{0.125} & 0.098 & \textbf{0.092} \\
case1 & 0.271 & \textbf{0.258} & 0.254 & \textbf{0.245} & 0.254 & \textbf{0.244} & 0.114 & \textbf{0.102} \\
case2 & 0.271 & \textbf{0.258} & 0.254 & \textbf{0.245} & 0.254 & \textbf{0.244} & 0.114 & \textbf{0.102} \\
case3 & 0.271 & \textbf{0.258} & 0.254 & \textbf{0.245} & 0.254 & \textbf{0.244} & 0.114 & \textbf{0.102} \\
case4 & 0.271 & \textbf{0.258} & 0.254 & \textbf{0.245} & 0.254 & \textbf{0.244} & 0.114 & \textbf{0.102} \\

\hline
\end{tabular}
\end{table*}

The results on the artificial networks with spammers are shown in Fig. 4. As expected, the RS in Fig. 4 are much higher than that in Fig. 3. This results suggest that the spamming behavior seriously decreases the ability of the ranking algorithms to detect the true quality of items. The interesting phenomenon is that, even with the spammers existing in the networks, the conclusion we drew from Fig. 3 still stands. That is, $p_1$ hardly influences the $RS$ while $p_2$ substantially affects the $RS$.

The detailed values of the $RS$ in Fig. 4 are reported in table 3. We also show the performance of the other two ranking algorithms, namely the mean and IR methods. The results of these two algorithms are also shown in table 3. In this table, the column called "projected" means that we use the projected rating as the input data to the ranking algorithms and we selected the optimal $p_1$ and $p_2$ parameters. Clearly, the $RS$ with the projected rating is significantly lower than that with the original rating, consistent in all ranking algorithms.

In all the results on artificial networks, it is shown that the ranking accuracy would achieve a maximum when $p_1$ and $p_2$ are set as either $0$ or $1$. This phenomenon is also observed in the simulation on real networks. These results indicate that 5 star rating is not effective for detecting high quality items. In this rating systems, the ranking of item quality mainly rely on rating 1 and 5. Based on this observation, we conjecture this might be the reason that "like-dislike" rating systems are so popular in many well-known websites such as \emph{youtube.com} and \emph{facebook.com}.



\section{Conclusion and discussion}
In this paper, we propose a rating projection method which allows us to project the original rating values to more proper ones. This method takes into account the nonlinearity between different rating values. To validate the method, we consider both artificial and real networks. The projected ratings are then applied to several ranking algorithms and significant improvement in the ranking accuracy is observed. Our simulations suggest that, to accurately rank the item quality, the detail rating information is not necessary. It is already sufficient if users can distinguish "like" or "dislike" an item.

Our work highlight the importance of the data pre-processing in ranking users' reputation and items' quality. Meanwhile, it raises a couple of unsolved problems. For example, the analysis in this paper is based on the 5-star rating systems. How to model and design projection methods for other rating systems such as 10-score rating systems and thumbing up systems still ask for further investigation. Moreover, the optimal selection of the parameters in the rating projection may vary in different real systems. A better way to determine the optimal parameters needs to be addressed in the future.

Finally, the idea of rating projection can be extend to a much wider context. Weighted network has always been a hot topic in network research. In many real network including neural networks and scientific collaboration networks, link weights play an significant role and usually used to represent the strength of the connections between nodes. Many works have devoted to design methods that can make best use of the weights for various objects such as link prediction, link salience and rich club detection. We believe that our method can also be applied to adjust the link weights, and similar improvement in these fields are expected.


\acknowledgements
The authors would like to thank Alexandre Vidmer for his valuable comments on this work. This work was partially supported by the EU FP7 Grant 611272 (project GROWTHCOM) and by the Swiss National Science Foundation (grant no.~200020-143272).

\end{document}